\documentclass[12pt]{iopart}
\usepackage{graphicx}
\usepackage{epstopdf}
\usepackage{dcolumn}
\usepackage{bm}
\usepackage{subfigure}
\usepackage{color}
\usepackage{lineno}
\usepackage{marginnote}
\usepackage{hyphenat}
\usepackage{indentfirst}

\usepackage{iopams}

\begin{document}

\title{Deformation of loops in 2D packing of flexible rods}

\author{T A Sobral$^{1,2}$, V H de Holanda$^{1,3}$, F C B Leal$^1$, T T Saraiva$^{1,4}$}

\address{$^1$ Departamento de F\'isica, Universidade Federal de Pernambuco, 50670-901, Recife, Brazil}

\address{$^2$ Instituto Federal de Educa\c{c}\~{a}o, Ci\^{e}ncia e Tecnologia do Rio Grande do Norte, RN 288, s/n, Nova Caic\'{o}, 59300-000 - Caic\'{o}, Brazil}

\address{$^3$ Instituto Federal de Educa\c{c}\~{a}o, Ci\^{e}ncia e Tecnologia do Sert\~{a}o Pernambucano, PE, Zona Rural, 56915-899 - Serra Talhada, Brazil}

\address{$^4$ National Research University Higher School of Economics, 101000, Moscow, Russia}

\ead{tteixeirasaraiva@hse.ru; thiago.sobral@ifrn.edu.br}
\vspace{10pt}
\begin{indented}
\item[]\today
\end{indented}

\begin{abstract}
The injection of a long flexible rod into a two-dimensional domain yields a complex pattern commonly studied through elasticity theory, packing analysis, and fractal geometries. ``Loop'' is a one-vertex entity that is naturally formed in this system.
The role of the elastic features of each loop in 2D packing has not yet been discussed.
In this work, we point out how the shape of a given loop in the complex structure allows estimating local deformations and forces.
First, we build sets of symmetric free loops and performed compression experiments.
Then, tight packing configurations are analyzed by using image processing.
We found that the dimensions of the loops, confined or not, obey the same dependence on the deformation.
The result is consistent with a simple model based on 2D elastic theory for filaments, where the rod adopts the shape of Euler's elasticas between its contact points.
The force and the stored energy are obtained from numerical integration of the analytic expressions.
In an additional experiment, we obtain that the compression force for deformed loops corroborates the theoretical findings.
The importance of the shape of the loop is discussed and we hope that the theoretical curves may allow statistical considerations in future investigations.
\end{abstract}

\vspace{2pc}
\noindent{\it Keywords}: elasticity, elastic loops, elastica, packing, pattern formation
\section{Introduction}\label{Intro}

The injection of filaments into cavities is a basic problem involving elasticity and self-exclusion.
These are two aspects of great importance in nature, with wide range of influence from polymeric packing~\cite{flory53,gennes79} to DNA packaging in viral capsids~\cite{gomes08,Stoop11, Vetter14,Holanda16}.
The structures formed by the confinement of wires packaged \cite{gomes08,mec_fio3D} and crumpled surfaces \cite{mec_CS} present anomalous characteristics \cite{gomes08} that are of interest to soft matter physics and statistical physics. 
The mechanical properties are influenced by the hierarchy of its folds~\cite{Donato02,Deboeuf09} and show a behavior similar to that found in unbranched polymers diluted in a solvent \cite{gennes79}.
We focus on the two-dimensional confinement of an elastic slender rod, which present interesting scaling properties~\cite{Donato03}, energy distributions~\cite{Boue07,Lin08, Deboeuf09}, and morphological phases~\cite{Stoop08}. The rigidity of the structure~\cite{gomes08} is of particular interest in the present study.

\begin{figure}[!ht]
\centering
\includegraphics[width=8.5cm]{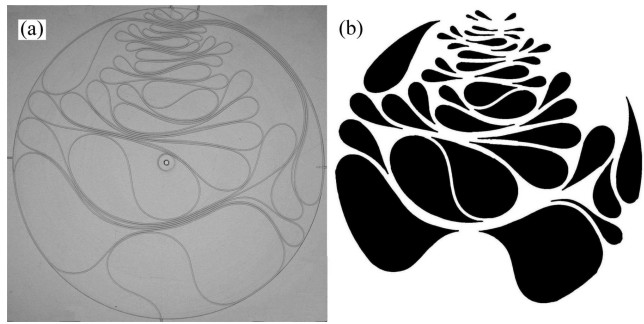}
\caption{\label{Fig1} (a) The packing of a 1~mm thick nylon fishing line into a circular cavity with a diameter of 200~mm. (b) The loops (in black) are closed domains with one vertex.}
\end{figure}

In a typical 2D confinement as shown in Fig.~\ref{Fig1}(a), the formation of self-contact points divides the area of the cavity into cells with a variable number of vertexes.
``Loop'' is the term used to designate single-vertex cells~\cite{Donato02}, as those highlighted in Fig.~\ref{Fig1}(b).
The determination of the loops allows identifying the points of higher curvatures~\cite{Donato07} and helps to determine the morphology of the complex pattern~\cite{Stoop08}.
The key concept here is \textit{curvature}, which links our work to the physics of pattern formation in protein chains and membranes~\cite{canalejo17} and, not less astonishingly, quantum gravity~\cite{cunha09}.
Besides, the number of loops gives the jamming length and the dynamic state of the system~\cite{Sobral15b, Sobral17}. Note that the type of loops studied here, in which curvature reverses sign along its length, can be unstable or less frequent in the case of 3D confinements, where torsion may have a major influence and the curvature maintains a constant sign~\cite{Morigaki16}.
Even so, racquet shapes were observed in fluctuating polymers~ \cite{Schnurr2000,Schnurr2002}, nanotubes and biofilaments~\cite{Cohen2003}, and the folding of paper stripes~\cite{Mahadevan1999}.
Here, we report a set of three experiments and a model that describes the deformations of the loop.
The main objective is identifying physical quantities in the complex confined system by analyzing the shape of a single loop.

The article is organized as follows: in Sec.~\ref{sec2}, we show our measurements of fundamental geometric quantities of single loops under compression and compare them to the case of loops from a packed conformation of longs rod inside 2D circular and rectangular cavities; in Sec.~\ref{sec3}, one can find solutions for the elasticity equation in 2D for the loop under compression; in Sec.~\ref{sec4}, it is described the experiment for measuring the force of compressing and stretching the loops in the configurations obtained in the previous section; and in Sec.~\ref{conc}, we show our conclusions.

\section{Packing and compression of elastic loops}\label{sec2}


The objective of the first experiment is to determine the mechanical response of a single loop under compression.  
The loops were constructed by bending a piece of rod to merge the ends into a vertex point. 
A bent rod leaves the planar configuration when the ends are close to each other~\cite{Goss05,Bosi15}, then transparent parallel plates are used to constraint it. 
The rods used are either nylon fishing lines or polymeric tubes. 
The fishing lines (packed in circular cavities) are 0.80 mm thick, and they are composed of nylon,  with Young's Modulus of 2.7 (8) GPa and a Poisson ratio of 0.39. 
The polymeric tubes (packed in rectangular cavities) are 5.0 (2.5) mm in external (internal) diameter, and they are composed of flexible polyvinyl chloride, with Young's Modulus of 3.3(1) GPa and the Poisson ratio of 0.38.
The constraint plates are made of acrylic with a thickness of 1.0 cm.
We also build planar loops with 1.0 cm wide ribbons and made of 0.10 mm thick A4 paper or 0.21 mm thick transparent Mylar sheets. 
An interesting point is that loops made from ribbons with large width are naturally planar~\cite{Morigaki16}, and have the advantage of eliminating friction effects due to contact with parallel plates.

\begin{figure}[!t]
\centering
\includegraphics[width=8.cm]{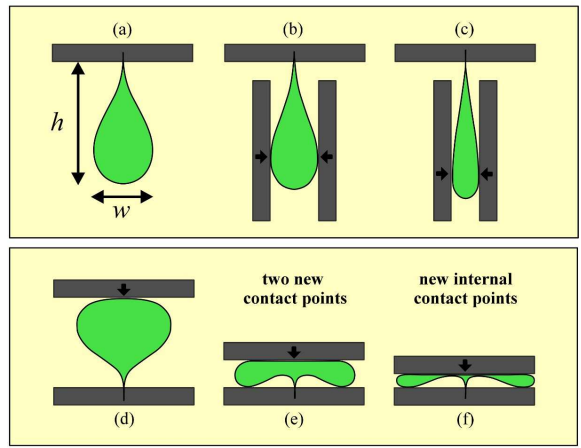}
\caption{\label{Fig2} Superior view of deformed loops: (a-c) compressing sideways by imposing smaller widths, and (d-f) compressing along the symmetric axis by imposing smaller heights.}
\end{figure}

The initial conformation is denominated ``free loop''.
The perimeter $\lambda=\lambda_0$ is defined when the flexible rod or ribbon is straight and was chosen randomly in an interval $\lambda_0 = \{97 - 366\}$~mm. 
The loops are large to avoid plasticity effects.
The height $h$ of the loop is defined along its symmetric axis as the distance from the vertex to the top of the bulge.
Perpendicularly, the largest width of the loop is $w$ [Fig.~\ref{Fig2}(a)].
Both $h$ and $w$ are measured with a digital caliper. 
The initial ratios $h_0/\lambda_0 = (0.424\pm0.007)$ and $w_0/\lambda_0 = (0.210\pm0.008)$ are essentially the same irrespective of the type of material. For rods, we measured the dimensions $h_0$ and $w_0$ from the neutral axis.

The experiment consists of compressing the loop along two perpendicular axes (Fig.~\ref{Fig2}). 
In the first part, the loop is compressed between two parallel aluminum plates, requiring $w$ to decrease as illustrated in Fig.~\ref{Fig2}(a-c).
Care is taken to preserve the loop as symmetric as possible. 
The system responds by increasing the height $h$. 
The compressing continues until the physical limit $w \rightarrow 0$ which leads to $h\rightarrow \lambda/2$. 
During this process, the contact points between the loop and the compression plates move downwards as indicated by small arrows in Fig.~\ref{Fig2}(b-c). 
In the second part of the experiment, the loop is compressed between two parallel aluminum plates, requiring $h$ to decrease as illustrated in Fig.~\ref{Fig2}(d-f). 
In this case, the system responds by increasing $w$. 
The compressing continues until the physical limit $h \rightarrow 0$ which leads to $w\rightarrow \lambda/2$. 
The experiment is performed horizontally over a table without gravitational effects.

\begin{figure}[!ht]
\centering
\includegraphics[width=8.5cm]{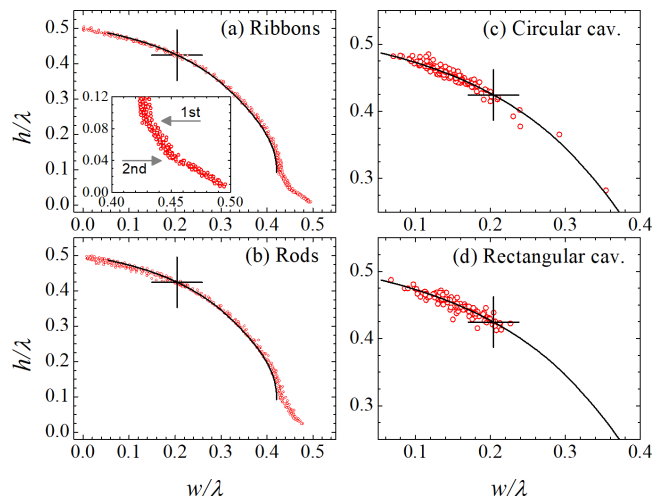}
\caption{\label{Fig3} The $h \times w$ diagram shows (I) the deformation of single loops made of (a) mylar ribbons and (b) fishing lines; and (II) image processing of packaged loops (c) in circular and (d) in rectangular cavities. The cross indicates the sizes of the free loop. The solid line indicates the theoretical curve. The inset in (a) shows a zoom to the lower right corner where corners appear due to the new contacts between the elastic line and the rigid plane. See text for further details.}
\end{figure}

The results of the experiment illustrated in Fig.~\ref{Fig2} are shown in Fig.~\ref{Fig3}(a-b) as a $h \times w$ diagram, normalized by $\lambda$. 
The diagram is identical for loops built from ribbons (Fig.~\ref{Fig3}a) and rods (Fig.~\ref{Fig3}b) and agrees with the theoretical description (solid line) based on Euler's Elastica which we discuss afterward. 
The values for the free loop are indicated by a big cross. 
Since compressing $w$ leads to higher $h$,  the graph in Fig.~\ref{Fig3}(a-b) must be read from the cross to the left. 
On the other hand, compressing $h$ leads to higher $w$, and the graph in Fig.~\ref{Fig3}(a-b)  must be read from the cross to the right. 
Therefore, the range of the data shown in Fig.~\ref{Fig3}(a-b) covers the entire physical domain, from values of dimensions for the free loop to completely crushed shapes.

Due to its anisotropic shape, the loop's response to compression depends on the direction.
As $w$ decreases, the curvature becomes localized at a single point, the tip of the loop, and the rest of the line is substantially straight, as depicted in Fig.~\ref{Fig2}(b-c).
As a consequence, in the limit $w \to 0$, the height $h \to \lambda/2$ and the diagram has no corners.
Compressing in the perpendicular direction, as shown in Fig.~\ref{Fig2}(d), a reduction in $h$ leads to an increase in $w$ until we reach the first corner point at $w_1/\lambda = (0.429\pm0.005)$, as displayed in Fig.~\ref{Fig2}(e). By decreasing $h$ further, the curvature is distributed sideways, and the new bulges eventually touch the compressing plane, as illustrated in Fig.~\ref{Fig2}(f). If the compression continues beyond this second point at $h_2/\lambda=(0.042\pm0.002)$, the curve follows towards the limiting value $w\to\lambda/2$.



\begin{figure}[!ht]
\centering
\includegraphics[width=7.cm]{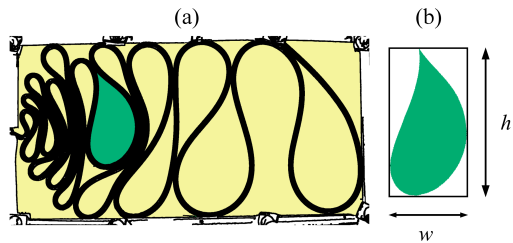}
\caption{\label{Fig4} (a) Digital image of the tight-packed flexible tube of $50$~mm in diameter inside a rectangular cavity of $400  \times 200$~mm$^2$. (b) Each loop is framed and rotated to maximize the height inside a rectangle. The process defines $h$ and $w$ (see text for details).}
\end{figure}

To study the problem of 2D confining, the flexible rod can be split at the contact points into pieces that are treatable numerically with Euler's equation~\cite{Boue06, Deboeuf09}. 
The fractal distribution of contact points along the wire in the cavity was studied in Ref.~\cite{Donato07}. Instead of considering the intricate propagation of forces along all the length of the wire~\cite{Donato07,Stoop08}, we proposed a simpler method based on the shape of loops in 2D cavities to obtain both qualitative and quantitative information about local deformations and energy.
The second experiment consists of the two-dimensional injection of thin rods into planar slender cavities. 
The objective is to analyze how the loops, which are naturally formed in the confinement, behave in light of the $w \times h$ diagram. 
This experiment is also divided into two parts. 
In the first one, nylon fishing lines with a diameter of $1.0$~mm are injected into circular cavities with $201$~mm in diameter as shown in Fig.~\ref{Fig1}(a). 
In the second part of the experiment, polymeric tubular rods with $50$~mm in diameter are injected into rectangular cavities of $400 \times 200$~mm$^2$ as illustrated in Fig.~\ref{Fig4}(a). To ensure two-dimensional confinement, the thickness of the cavities allows only one layer of the rod.
All injections are performed manually into a dry cavity, free of lubricants, at a rate of about $1$~cm/s. 
The injection stops when the system jams due to the rigidity around the injection channel.

The particular conformation presented in Fig.~\ref{Fig4}(a) is chosen as a representative case where we can briefly discuss the role of elasticity, self-exclusion, and friction with the cavity. 
The first loops that formed are pushed away from the injection channel, shown in the right region of Fig.~\ref{Fig4}(a).
These loops are larger and require low values for the injection force in the early stages.
Observe that the shapes of the loops are prevented from changing due to the contacts with the cavity and friction.
The rod has small self-contact regions and the stacked loops transmit the force through the system by compressing each other.  
The geometric pattern and local rigidity are then governed by the elasticity of the loops. 
On the other hand, the last loops are close to the injection channel, see the left region of Fig.~\ref{Fig4}(a).
The injection force in this stage is higher, which reduces the perimeter of the loops and limits them to occupy the periphery of the pattern, without contacting the cavity.
The rod has large self-contact regions that transmit the force through the system.
In the case shown in Fig.~\ref{Fig4}(a) the loops  are compressed by neighbors in a configuration that resembles Fig.~\ref{Fig2}(a-c). 
In a cavity of generic shape, however, we expect that the loops can interact also as in Fig.~\ref{Fig2}(d-f). 
For example, there are differences between patterns generated from circular and rectangular cavities [Compare Fig.~\ref{Fig1}(a) and Fig.~\ref{Fig4}(a)].

To build the $h \times w$ diagram for loops within cavities, we need to measure their dimensions non-invasively. Digital photos were taken with an Olympus C-3040ZOOM digital camera, positioned over the transparent cavity on a white horizontal table. The image is subjected to some filtering operations, edge detection, and morphological components using generic image processing programs, which allow selecting the loops, as shown in Fig.~\ref{Fig4}(a). Within the cavities, loops interact intricately and assume non-symmetrical shapes, so we redefine $h$ and $w$ appropriately for this context.
We focus here on the reading of the geometric pattern by prioritizing simple definitions \footnote{For readers interested in numerical methods to quantify the shape of planar curves, see also Cartan’s invariant signature \cite{Calabi1998}. Note that both methods are invariant under Euclidean transformations.}.
First, the selected loop is inscribed into a rectangle. 
Rotating the loop will change the aspect ratio of the frame. 
The chosen angle is that which maximizes the height of the frame, as shown in Fig.~\ref{Fig4}(b). 
This dimension is identified as $h$ or $w$ accordingly to the position of the tip and the bulge of the loop.
Following this procedure, all symmetrical loops illustrated in Fig.~\ref{Fig2} maintain their dimensions.
The points at the edge of the loop are interpolated by a curve that allows us to measure the dimensions $h$, $w$, and $\lambda$ with the same units.
Such a method is repeated for each loop in the image. 
The result is shown in Fig.~\ref{Fig3}(c) for loops inside circular cavities, and in Fig.~\ref{Fig3}(d) for loops inside rectangular cavities. 

The main result in Fig.~\ref{Fig3} is that the loops obtained from the packing of rods in a 2D cavity, whether in a circular or rectangular cavity as shown in Figs.~\ref{Fig3}(c-d), follow the same diagram $h \times w$ as single handmade loops from mylar ribbons and fishing lines, shown in Figs.~\ref{Fig3}(a-b).
However, the data from packed loops distribute themselves mostly on the left side of the diagram, instead of spreading over the whole diagram. 
This is in agreement with the fact that the loops are aligned perpendicular to the injection channel. 
The injection force then acts to compress the loops laterally, reducing both their perimeter and width.
However, Fig.~\ref{Fig3}(c) shows that few loops are also longitudinally compressed in circular cavities. 
In this cavity rotation and slippage are more accessible than in the rectangular one.
In the general sense, this confirms that how the loops populate the diagram depends on the shape and size of the cavity. 

\section{Theoretical description}\label{sec3}

\begin{figure}[!ht]
\centering
\includegraphics[width=8.5cm]{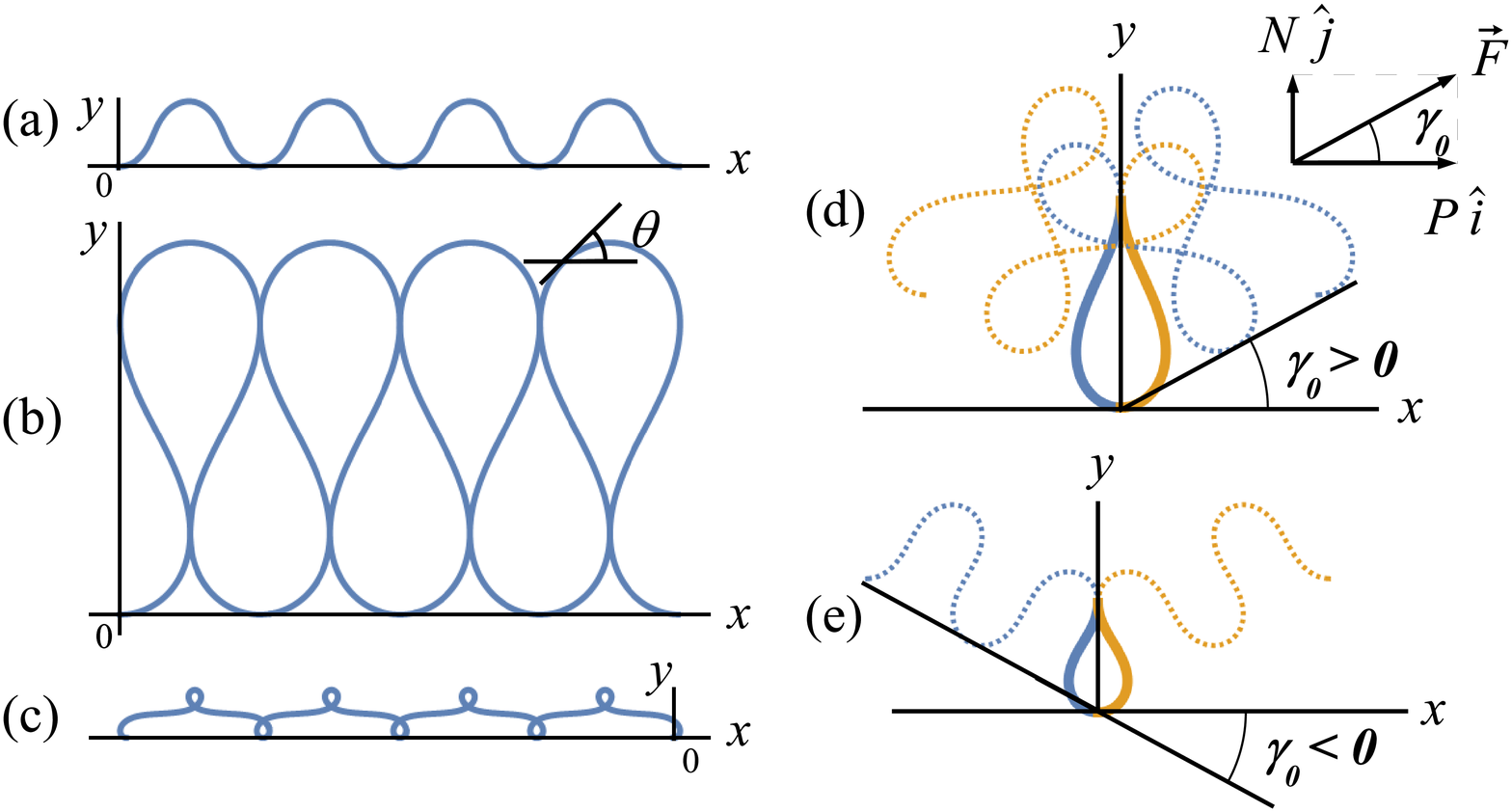}
\caption{\label{Fig5}
The family of solutions of Eq.~\ref{Eq1} for $x(0) = y(0) = \theta(0) = 0$ is shown for (a) $p = 0.3$, (b) $p^* = 0.73118$, and (c) $p = 0.999$. Here, the deformed loop is modeled by a solution in coordinates rotated by an angle (d) $\gamma_0=\pi/6$~rad ($p^*=0.906799$) and (e) $\gamma_0=-\pi/6$~rad ($p^*=0.509152$) and its reflection on the vertical axis.
Note that the solutions are periodic along the line of force.
}
\end{figure}

The theoretical description follows the elastic model for filaments, which allows determining the shape of single loops~\cite{Landau86,Nizette99,Goss05}.
Besides, values of compressive force and elastic energy stored in the loops are found as functions of deformation.
We describe the filament by a purely elastic curve. The force vector is constant along the slender rod, and it points in the direction of a horizontal x-axis, $\mathbf{F}(s) = F \, \hat \mathbf{i}$. 
Since $\theta$ is the angle between the infinitesimal length $d \mathbf{s}$ and the direction of the force $\mathbf{F}$, the shape is governed by infinitesimal torques $d\Gamma(s)=  F \sin{\theta} \,ds$. 
The elasticity of the rod is given by the bending rigidity $\mu$, which relates the torque to the local curvature, $\Gamma(s) = - \mu \frac{d \theta}{d s}$.
From this, we can write a second-order differential equation for the angle:
\begin{equation}
\frac{d^2 \theta}{d \tilde{s}^2} = - \sin{\theta},
\label{Eq1}
\end{equation}
where the tilde over any quantity $q$ means $\tilde{q} \equiv q \, \sqrt{F/\mu}$. 
We chose the initial angle $\theta(0) = 0$ and the initial curvature $\dot \theta_0$ (used as a parameter) as boundary conditions.
The Cartesian's coordinates are given by integration of $d \tilde x(\tilde s) / d \tilde s = \cos{\theta}$ and $d \tilde y(\tilde s) / d \tilde s = \sin{\theta}$, starting from the origin: $\tilde x(0) = \tilde y(0) = 0$.
The family of solutions [Figs.~\ref{Fig5}(a-c)] are written in terms of elliptic integrals and elliptic functions, and the elliptic parameter $p \equiv \dot \theta_0^2/4$ determines the overall shape~\cite{Nizette99}.

The free loop, sketched in Fig.~2(a), presents a special point where the branches of the rod tangentially merge.
Therefore, we selected the solution with a single self-contact point, at $\tilde s = \tilde s^*$, by imposing the conditions:
\begin{equation}
\tilde x(\tilde s^*) = \tilde x(0) \textrm{\hspace{10pt} and \hspace{10pt}} \frac{d \tilde x(\tilde s)}{d \tilde s} \Bigr|_{\tilde s=\tilde s^*} = 0.
\label{Eq2}
\end{equation} 
That allows us to write an equation for $p$ in terms of elliptic functions. 
The numerical solution provides the known value $p^*=0.731183$~[22].
Regarding the dimensions of the free loop, the length $\tilde s^*$ characterizes a half perimeter $\tilde \lambda/2$, the coordinate $\tilde y$ of the vertex point defines the height $\tilde h$, and the maximum value of $\tilde x$ delineates the half-width $ \tilde w/2$ of the free loop.
We find $h_0/\lambda = 0.424308$ and $w_0/\lambda = 0.204214$, irrespective of the bending rigidity. 
Those values agree with the experimental finding, as the big crosses in Fig.~3 shows. Once we find the value of $\tilde s^*$, the force required to build the free loop is $F = \mu (\tilde s^*/s^*)^2 = 4 \mu (\tilde s^* / \lambda)^2$.

Our model for a deformed symmetrical loop requires attention to the contact points, where external forces act. As we can see in Fig.~\ref{Fig2}, there is a contact point in the middle of the loop when we compress it longitudinally and two contact points when we compress it laterally. 
To simplify the problem, we assume that lateral compression of the loop is equivalent to longitudinal elongation.
In this manner, all compression in Fig.~\ref{Fig2} can be described by the application of a single normal force $\mathbf N = N \hat{\mathbf{j}}$ at the middle point of the loop, with $N>0$ for the compressing sideways and $N<0$ for longitudinal compressing. 
This approach proved to be sufficiently accurate when the results corroborate the experimental data well.

For the deformed loop, the vertical force changes the reference axis of the curve, because the total force has now two components $\mathbf{F} = P \hat{\mathbf{i}} + N \hat{\mathbf{j}}$ [Fig.~\ref{Fig5}(d)]. 
Eq.~(\ref{Eq1}) remains valid for a coordinate system rotated by an angle of $\gamma_0 \equiv \tan^{-1} \left( N/P \right)$, but the contour conditions, Eq.~(\ref{Eq2}) are the same.
That leads us to write an equation for $p$ in terms of elliptical functions, as before, with the solution $p^*$ now depending on the angle $\gamma_0$.
Therefore, the deformed loop is composed of two mirrored elasticas that start from the origin and merge at the vertex, as can be seen in Fig.~\ref{Fig5}(d-e).

\begin{figure}[!ht]
\centering
\includegraphics[width=8.5cm]{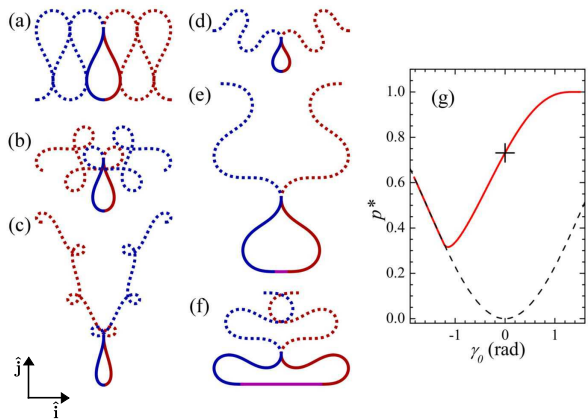}
\caption{\label{Fig6} 
The symmetric loop is composed of two elasticas (a-f) rotated respectively by $\gamma_0 = \{$0, 0.60, 1.20, -0.55, -1.27, -1.82$\}$~rad. 
(g) $p^*$ dependence with the compression force through $\gamma_0$.
The dashed line corresponds to $\dot \gamma_0=0$ case. See text for details.
}
\end{figure}

Figure~\ref{Fig6}(a-f) shows some situations where two symmetrical elasticas (dashed lines) merge to compose the deformed loop (solid lines). 
The free loop corresponds to $\gamma_0=0$, illustrated in Fig.~\ref{Fig6}(a).  
In the case $\gamma_0>0$, shown in Fig.~\ref{Fig6}(b-c), the curve intersects, indicating that the solution $p^*$ is greater than that of the free loop.
In the case $\gamma_0<0$, shown in Fig.~\ref{Fig6}(d-f), the curve no longer intersects, indicating that the solution $p^*$ is less than that of the free loop.

Figure~\ref{Fig6}(g) summarizes our results by showing the dependence of the elliptical parameter $p^*$ as a function of the angle $\gamma_0$.
As usual, the solution for the free loop is illustrated by the large cross.
In general, the elliptical parameter $p^*$ departs from the free loop result for $\gamma_0 \neq 0$.
The limit case $\gamma_0 \to \infty$ leads to the homoclinic elastica $p^* \to 1$. At this limit, the loop is mostly a straight line with a very high curvature concentrated in a minuscule region.
On the other hand, when $\gamma_0 < 0$, $p^*(\gamma_0)$ is not a monotonic function because of a physical constraint. 
The dashed line in Fig.~\ref{Fig6}(g) shows the values of $p^*$ that satisfy the zero curvature condition at the origin.
This condition in the rotated coordinate system provides $p^*=\sin^2(\gamma_0/2)$. 
That is a physical constraint in our experiment because the compression plate does not allow for a change in the curvature signal, as Fig.~\ref{Fig2}(e) shows well.
To extend our model somewhat beyond this limitation, we have included a straight line between the origin and the inflection point, as illustrated in Fig.~\ref{Fig6}(f).
We do not cover the entire region $ \gamma_0 < 0$ because new self contacts appear when the deformation is high enough [Fig.~\ref{Fig2}(f)].

Once the parameter $p^*$ is found, the appropriate planar elastica is defined.
Although the problem is solved in rotated coordinates, we define the central quantities of interest in the original coordinate system: the half perimeter $\tilde \lambda/2 = \tilde s^*$ obeys the condition $\tilde x(\tilde s \neq 0) = 0$ (length $\tilde s^*$ of the vertex), the height $\tilde h$ is given by the coordinate $\tilde y(\tilde s^*)$ and the width $\tilde w$ is obtained by the maximum value of $\tilde x(s)$ between 0 and $ s^*$.
In the extension of the model for $\gamma_0<0$, only the half perimeter needs adjustment to $\tilde \lambda/2 = \tilde s^* + \tilde \ell$, where $\tilde \ell$ is the distance between the origin and the inflection point (fixed at $s = 0$).

One builds a parametric $h(\gamma_0) \times w(\gamma_0)$ diagram. Fig.~3 shows the result as a continuous line. Despite the approximations and extensions that we added to a purely elastic model, the diagram describes very well the general behavior of deformed loops even in the complex case of two-dimensional packing.
That inspired us to perform a third experiment, exploring the fact that the force profile  $F(\gamma_0) = 4 \mu (\tilde s^*(\gamma_0) / \lambda)^2$ is theoretically available.


\section{Measuring the force for compression and stretching of loops}\label{sec4}

\begin{figure}[!thb]
\centering
\includegraphics[width=0.6\linewidth]{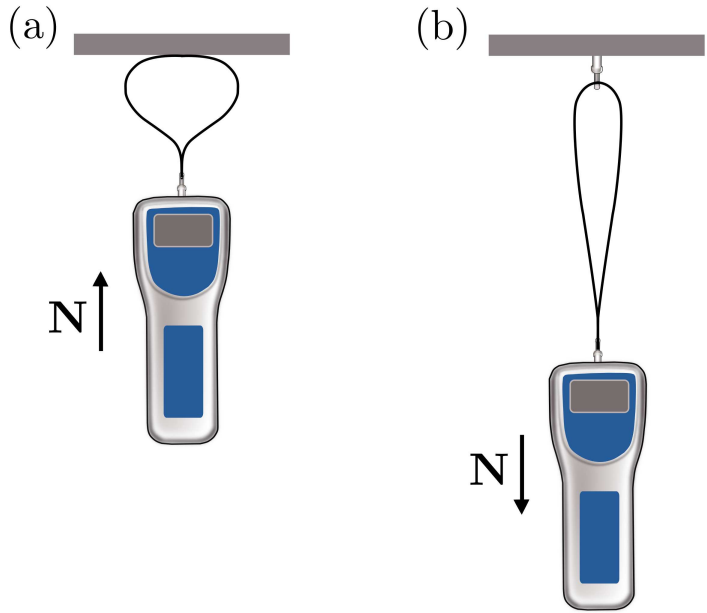}
\caption{\label{Fig7} Schemes of the measurements of the force N (a) to compress the loop and (b) to stretch the loop. This experiment was executed along the horizontal plane perpendicular to the gravitational force.}
\end{figure}

In a third experiment, we measured the force $N$ versus deformation ($\Delta h / \lambda$) through an Instrutherm DD-500 dynamometer, with instrumental precision of 0.01 N, coupled in automatic equipment, whose spatial displacements were given in steps of 2 mm. The loop samples were made with strips of acetate sheets 0.5 mm thick and $1.00 \pm 0.02$ cm wide. 10 measurements were made with strips of different sizes. The measurements obey two steps: (i) we attach the vertex of the loop to the probe tip of a dynamometer and (ii) we reset the dynamometer measurement after connecting the loop to it. We compress the loop by pushing the protrusion against a fixed wall (see Fig.~\ref{Fig7} (a) - in this process $N < 0$), while in the stretch we fix a metal hook on its protuberance, then we stretch the loop by pulling the dynamometer (see Fig.~\ref{Fig7} (b) - in this process $N > 0$).

\begin{figure}[!thb]
\centering
\includegraphics[width=7.5cm]{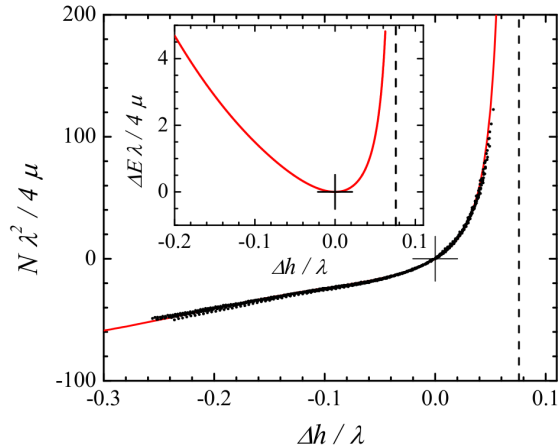}
\caption{\label{Fig8}  The force $N$ as a function of the deformation $ \Delta h / \lambda $ for elastic loops. The black dots correspond to the data obtained with the experimental setup. The continuous red line corresponds to the theoretical prediction, while the dashed lines indicate the physical limit of deformation. The inset shows the elastic energy stored in the loops, $ \Delta E $, as a function of the deformation $ \Delta h / \lambda $. See text for details.}
\end{figure}

The results obtained for the measurements of the force (N) were shown in Fig.~\ref{Fig8} as black dots. The force increases faster during the stretching than in the compression process due to the asymmetry of the loop. The dashed line in Figure~\ref{Fig8} points to the limit h tending to $\lambda/2$, where the force and energy diverge because of the small radius of curvature of the strip due to stretching. The agreement between the model and the experiment allows us to estimate the strength and energy stored for loops within complex patterns.


The agreement between the theoretical curve and the experimental data (Fig.~\ref{Fig8}) shows the relevance of this purely elastic model. We can calculate the force associated with a given shape, $F=\mu(\tilde{s},s)^{2}$. The result for the force is shown in Fig.~\ref{Fig8} and (inset) the energy cost of deformation is obtained by integrating the force $N$ over infinitesimal displacements of the height $\Delta E = W = \int N\ \mbox{d}h$.
Note that big crosses in Fig.~\ref{Fig8} indicate the values of the initial conditions.

Measurements of forces involved in the deformation of loops are useful to obtain a better understanding of the configurations of the injected wire inside the cavity. It is possible to integrate the force along the wire to obtain total energy in a specific configuration. On the other hand, it is possible to determine the energy distribution in the cavity using the loop density and our local energy measurements for each loop. We hope to carry out new experiments to test our hypothesis and theoretical model and to compare them with other previous publications~\cite{Donato07}. Also, we believe that our experimental and theoretical results may be useful in the study of other chain-like packing systems, such as the case of genetic material into virions. For example, these configurations are important in bacteriophages that need high pressures within the cavity to eject their genetic material to pass through the cell membrane~\cite{purohit01}.




\section{Conclusion}\label{conc}
In this work, we suggest new methods to analyze the deformation of a flexible rod packed into a 2D cavity.  The $h\times w$ diagram in Fig.~\ref{Fig3} enables us to calculate the perimeter of the loops and their deformation level, compared to a simple model based on Euler's Elastica~\cite{Nizette99}. The purely elastic case is an important reference~\cite{Boue07,Deboeuf13}, and the graphs in Fig.~\ref{Fig8} allow us to estimate the forces and stored energy in any loop immersed in a complex pattern.  In this physical system, friction and plasticity play very important roles~\cite{Stoop08,Lin08}, but the effect on each loop needs further attention.  Finally, the results of the dependence of the stored energy on the loop deformation open up new possibilities for the application of statistical techniques to 2D confining of flexible rods.



\ack
We acknowledge Professor Eduardo O. Dias from UFPE for fruitful discussions.
The present work was supported by CNPq, Conselho Nacional de Desenvolvimento Cient\'ifico e Tecnol\'ogico, Brazil (numbers 157218/2012-0, 141813/2016-4, and 152053/2016-6).


\section*{References}
\bibliographystyle{unsrt}
\bibliography{references}

\end{document}